# Atom-Wide Co Wires on Cu(775) at Room Temperature

Nader Zaki[1], Denis Potapenko[1], Peter D. Johnson[2], and Richard M. Osgood[1]

[1]Columbia University, New York, NY 10027

[2]Brookhaven National Labs, Brookhaven, NY 11973

We report on a new surface phase of the Co-vicinal-Cu(111) system which exhibits self-assembled uniform Co quantum wires that are stable at 300K. STM-imaging measurements show that wires will self-assemble within a narrow range of Co coverage and, within this range, the wires increase in length as coverage is increased. The STM images show that the wires form along the leading edge of the step rise, differentiating it from previously theoretically predicted atomic-wire phases. The formation of relatively long laterally un-encapsulated one- and two-atom wires also differentiates it from past experimentally observed step-island formation. Furthermore, our experiments also show directly that the Co wires coexist with another Co phase that had been previously predicted for growth on Cu(111). Our observations allow us to comment on the formation kinetics of the atomic-wire phase and on the fit of our data to a recently developed lattice-gas model.

## I. Introduction

One-dimensional nanoscale systems, including atomic chains or wires, have been predicted to display a wide range of unusual physical properties [1]. The compelling physics of low-dimensional phenomena has led to the exploration of atomic-wire preparation techniques. One approach has been the use of stepped surfaces to form regular arrays via self-assembly [2]. These large-area step arrays are suitable for photoemission studies of the electronic structure [3, 4]; furthermore, this approach allows the degree of coupling between wires to be changed controllably through choice of vicinal cut. However, in step-edge growth, more complex processes may occur, including growth of islands or interdiffusion into the underlying substrate. These processes lead to important unresolved questions regarding step-mediated self-assembly of bimetallic-wire arrays, including an understanding of the basic physical parameters, which can, in turn, guide the choice of a particular materials system and lead to growth methods that ameliorate chemical mixing.

Due to the interest in spin-valves, the bimetallic system Co/Cu(111) has been studied extensively [5-9]. Several experiments employed non-vicinal, i.e. with terrace widths ≥100Å, Cu(111) surfaces for spin-polarized investigations of Co nanostructures, most notably of 2ML high triangular islands [9]. However, the exact atomic make-up of these nanostructures remains in question. For example, it has been argued that the islands form on a buried layer of Co and that Cu migrates to the outer perimeter of the islands [7]. In one instance, room-temperature (RT) scanning tunneling microscopy (STM) has been used to examine the self-assembly of Co island chains, ~50Å in width, along the edges of isolated Cu steps [6]. These chains appeared as 4Å-high protrusions at the step edge and thus were neither single-atom in width or height. Finally, Monte Carlo simulations [10] and density functional theory (DFT) calculations [11] have been used to investigate the self-assembly mechanism of Co wires at a Cu(111) step. Though there are differences in the details of their predictions of an atomic-wire surface phase, both suggested that laterally encapsulated Co wires are formed during self-assembly.

In general, these earlier large-terrace studies show clearly that Co accumulates at the Cu(111) step edges during growth. However several compelling and important questions remain about surface self-assembly on narrow-terrace width vicinal Cu(111): first, can uniform Co wires of atomic width form at straight step edges; second, will such Co wires be laterally encapsulated by Cu at the step edge, as predicted theoretically; and finally, do other more complex Co atomic structures exist at RT if the terraces approach atomic dimensions.

In this paper, we answer these questions by using *in situ* STM as a probe of Co self-assembly on a Cu(775) step template. This substrate represents an $8.5°$ miscut of a Cu(111) surface, resulting in a terrace width of only 14.3Å or approximately seven atomic rows, separated by B-type steps. The crystallographic direction along the step edge is $[1\bar{1}0]$ while that perpendicular to the step edge is $[11\bar{2}]$. Our study concentrates on low-coverage regimes in order to observe and clearly identify the initial Co step-edge nucleation structure. As will be shown, our experimental findings clearly show a new phase at which low-coverage growth leads to self-assembly of long straight atomic wires.

## II. Co Nucleation On Cu(111) Steps: Prior Experimental Work

As mentioned in the introduction, there have been several experimental studies of Co deposited on Cu(111). One of the earliest studies, by Figuera *et al.* [12], looked at RT deposition of Co. In addition to bilayer triangular islands on terraces as well as 1ML terrace vacancies, they observed island nucleation at both the top and bottom of the step edge. This group pointed out subsequently [13] that the islands are comprised of a mixed phase of Co and Cu. Note that this morphology was observed at coverages as low as 0.1ML [13]. Finally note that a recent Co/Cu(111) study by Chang *et al.* [8] reported 2nm wide islands on the upper step edge at a coverage of 0.09ML.

Based on the above mentioned step nucleation morphology that is present for RT deposition, Figuera *et al.* [5] subsequently reported growth of chains of Co islands on Cu(111) step bunches. For a coverage of 0.2ML and an average step width of 10nm, the average island width was 5nm. Note that vacancy islands had also formed as a result of the step-island nucleation.

In addition to Co growth at RT, other groups have examined low-temperature deposition on Cu(111). For example, Pedersen *et al.* [7] observed "ramified islands", i.e. bilayer islands connected by monolayer islands, for Co deposited at 150K and imaged at 170K (i.e. not annealed to RT). The bilayer islands were found both at the bottom and top of the step edges, as well as on the terraces. This group also suggested that the bilayer islands sat, in fact, on an embedded layer of Co and that the islands were terminated on their perimeter by Cu atoms. Furthermore, this work showed that, if annealed to RT, interdiffusion of this Cu brim occurred. In a second low-temperature-deposition experiment, Speller *et al.* [14] deposited Co at 140K and then slowly elevated the sample to RT for imaging. At a coverage of ~0.12ML and a step density of 1/200 to 1/1000Å, they observed step island growth only at the top of the steps; however, step-island nucleation at the bottom of the steps commenced at about 0.4ML for their conditions. In their experiment, the islands were found to be a mixed surface phase of Co and Cu with an average width of 5nm. The intermixed nature of the islands was supported by the concomitant formation of vacancy islands.

In summary, despite extensive prior work in the Co/Cu materials systems, no atomic wires of Co on Cu(111) surfaces were observed experimentally, although the existence of Co wires at Cu step edges was predicted theoretically[10, 11]. Note, however, that there are several reports of the direct observation of atomic wires in different bi-metallic systems, e.g. Fe/Cu(111) [15].

## III. Experiment

Cu(775) was prepared using repeated sputter/anneal cycles in an UHV chamber with a base pressure less than $1\times10^{-10}$ Torr. Auger measurements showed no detectable contamination and surface electron diffraction showed the split spots characteristic of a stepped surface. Co was deposited from a heated ceramic crucible with the Cu sample cooled to ~165K. The cooling was a precaution against interdiffusion during sample preparation. Co coverage was measured using STM scans and assumptions of Co locations, which gave an estimated error of ~0.02ML. Using the aforementioned coverage estimation, the calculated average deposition rate for this work was 0.01ML/min. After deposition, the sample was brought to RT for the STM studies. All the scans were performed using a tungsten tip, the sample biased at +0.5V with respect to the tip, and a feedback tunneling current of 150pA.

## IV. Results and Discussion

An STM image of the clean Cu(775) surface is shown in Fig. 1. At room temperature, this surface exhibits unstable step edges or "frizz" [16] as shown in the atomic resolution image (Fig. 1(b)). This step instability has been attributed to kink motion at the step-edge or to step-tip interaction [17]. The steps in Fig. 1 have an average height of 2.1Å and the step edges appear rounded in both plan view and in profile due to tip-step convolution. As will be shown below, the lack of frizz denotes the presence of Co, which pins step-edge motion.

Figure 2(a) displays a derivative image taken at a Co coverage of ~0.09ML. Compared with the frizz at a clean Cu step, visible on the left, the figure shows that one-atom-wide wires have formed, distinguishable by their straight edges and uniform widths. The sharp or straight-edge appearance of the step-edge-assembled wires is attributed to the pinning of substrate step atoms by Co atoms. The measured "cross-section" over two steps indicated in the figure and shown in the inset suggests that the wires are approximately 3.0Å wide, which is comparable to the Cu row-row spacing of 2.2Å. Again, due to tip-step convolution, the step profile appears rounded resulting in a larger apparent wire width. Tunneling into the Co wires at a bias of +0.5eV above the Fermi level at constant current requires the tip to remain closer to the surface than for tunneling into the Cu substrate. Chemical contrast by STM is a well-known phenomena; see for example the related case of Co/Pt(111) [18, 19], which reports similar results to ours for both Co in the Pt terrace and at the step edge. However, though there have been several previous STM studies of the Co/Cu(111) system, this particular example of chemical contrast on this specific system, which allows the differentiation of Co step-edge wires from the Cu atoms on a Cu(111) terrace, has apparently not been commented on before. We note that a similar, though smaller effect, was observed for 2-3ML high Co islands surrounded by rims of several Cu atoms [7]. For one-atom wires this effect produces a small but clearly observable inflection in the step edge located approximately $1/2$ to $3/4$ up the step height of Cu. Note that this uniform inflection along the step edge is not seen at Cu steps where the step edge exhibits frizz; hence, this inflection is not likely due to a double-tip.

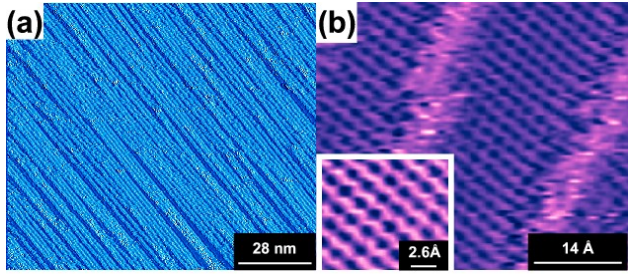

FIG. 1. (Color Online) STM images of Cu(775) at RT, taken at $V_{sample}$=0.5V and I=150pA; the images are *dz/dx* (i.e. derivative). (a) $100\times87nm^2$ scan showing relative uniformity of step spacing. (b) Atomic-resolution scan of steps showing "frizz". Inset, topographical close-up of step terrace clearly revealing the (111) surface.

As shown in Fig. 2(b), two-atom-wide wires were observed at a slightly higher Co coverage of about 0.12ML. These ~5.4Å-wide wires possess an average height of 1.5Å, less than that of a pure Cu step height, as seen for frizzy steps. This result further supports the discussion regarding the inflection measured in the profile of the one-atom-wide wires. We note that the local density of states may be affected by quantum-size effects due to the atomic scale of the wires, as seen, for example, in the case of atomically fabricated chains of Cu on Cu(111) [20].

The growth physics of these Co wires can also be examined via wire-length distributions measured from the STM images. Thus, in Fig. 3, one-atom wire histograms are given for the two different coverages mentioned above, namely 0.09ML and 0.12ML. The histograms clearly reveal an increase in the length of the wires as coverage is increased. Specifically, the average wire length at 0.09ML is found to be ~27 atoms with a standard deviation of 22 atoms and a maximum observed length of 105 atoms; at the higher coverage of 0.12ML, the average length had increased to ~40 atoms with a corresponding increased standard deviation of 37 atoms and a maximum observed length of 141 atoms. Note that the measurements are for one-atom wires only and do not include the wire-length distribution for 2-atom wide wires nor the terrace-site-exchange (see below), which are also observed at the higher 0.12ML coverage; hence, the one-atom wide wires increase in length despite the fact that these other atomic morphologies are starting to form. This observation of increasing wire-length with coverage provides insight into the wire self-assembly mechanism, as will be discussed below.

In addition to Co wires at the step edges, depressions are seen in step terraces, such as those in the regions marked by circles in Fig. 2(d). Several observations are consistent with these depressions reflecting Co atoms locally embedded in the terrace. First, the maximum apparent depth of these depressions is ~0.5Å which is consistent with the observed height difference between the wires and adjacent terraces. Second, this depth agrees well with similar 0.6Å deep depressions observed on the Co/Cu(001) system [21]. Third, if the apparent depressions happened instead to be terrace vacancies, i.e. 2Å deep, tip-convolution effects would likely not account for the measured shallow depth, since smaller features (2-atom wide wires) were also resolved along the same scan line of the depressions. Co-induced atom vacancies (etch pits) have been previously reported for Co/Cu(111) systems [5]. However, the pits, which were observed at a comparable Co coverage of 0.1ML, were much larger, ~2Å deep and 80Å in diameter, than found in the present study. Furthermore, the etch pits observed in the earlier study were commensurate with adatom-island formation, which is not seen at the same coverage in this study. Thus, as in the case for the wires, we attribute the depressions to a change in the local density of states (LDOS) due to Co–Cu substitution rather than a vacancy formation. We also note that this observation supports the long-held hypothesis [7] that Co/Cu(111) islands, seen by us at higher coverage, do indeed bind to a terrace-embedded Co layer. Given the narrow step width, the displaced Cu atoms most likely attach to the step edges and become part of a mobile step kink, appearing as frizz. For completeness, we note that Co is also known to adsorb on Cu(111) terraces as demonstrated by several low-temperature STM studies [22-24]; our observations show, however, that the temperature range we are operating in is energetic enough to allow for terrace-site exchange in addition to terrace diffusion.

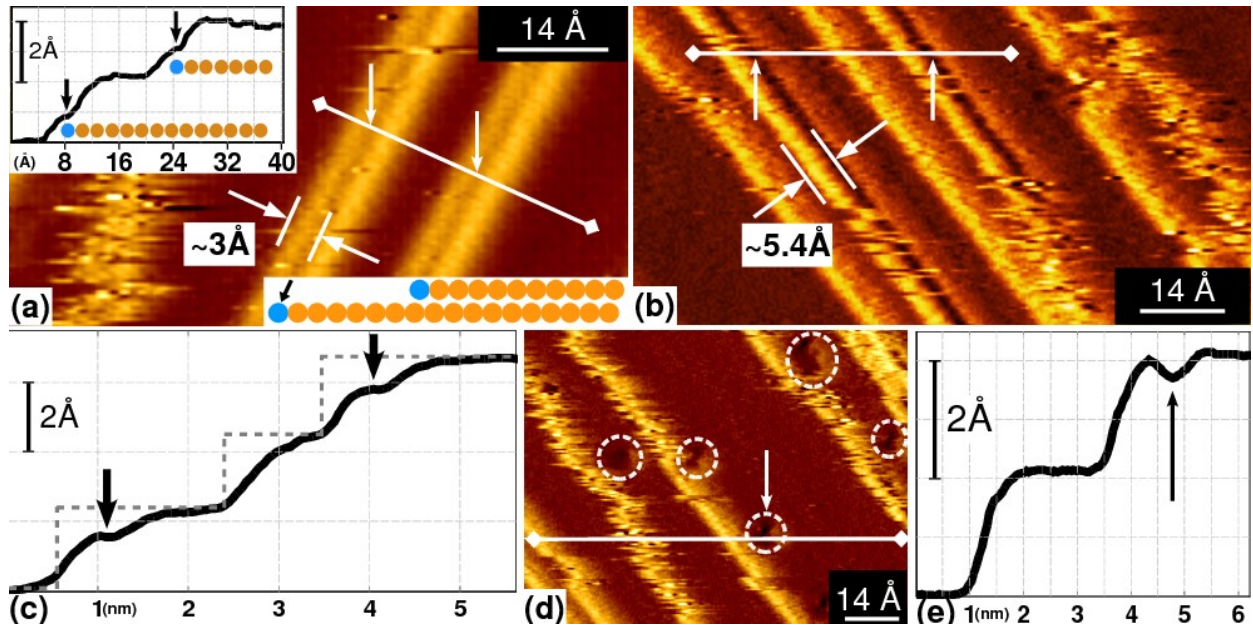

FIG. 2. (Color Online) STM derivative images of Cu(775) at RT after deposition of Co. Blue-colored circles denote Co atoms and gold-colored circles denote Cu atoms. The black arrows in the topographical profiles correspond to positions indicated by white arrows in the images. (a) Scan showing one-atom-wide Co wires; frizz is seen on a clean step edge at the left of the image. The inset is a topographical profile along the direction of the corresponding white line in the image. Coverage is 0.09ML. (b) Scan showing two-atom-wide Co wires, as well as one-atom-wide wires and clean-steps. A plateau, denoted by arrows, is seen for the two-atom wide wire scan; in this image, the plateau is not seen for one-atom-wide wires presumably due to finite tip radius and the narrowness of the wires; however, the lack of frizz is a strong indication of Co at the step edge. Coverage is 0.12ML. (c) Profile for white line in (b). The gray dashed line serves merely as a step guide for the eye and does not denote the exact position of the steps. (d) Scan showing terrace-embedded Co. Coverage is 0.12ML. (e) Embedded Co appears as depressions, as indicated by the arrow, in the step terrace.

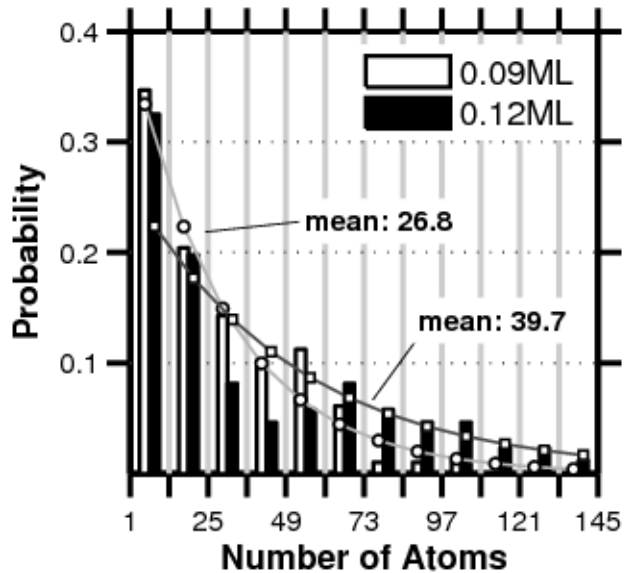

FIG. 3. Distribution of 1-atom-wide Co wire lengths. The smooth grey curves denote a fit based on a one-dimensional lattice gas model; see discussion in text. At a coverage of 0.09ML (circles), the average wire length is a ~27 atoms. As the coverage is increased to 0.12ML (squares), the average wire length increases to ~40 atoms, indicating that wire length increases with coverage. This result is in accord with a growth mechanism, in which the Co atoms move facilely along the step edge until encountering a wire end, to which the Co atom attaches. Note that in comparison to the typically reported widths of nanometer-scale islands at step edges (5nm or ~20 atoms), the 1- and 2-atom (not shown above) wires have grown to much longer length scales.

The occurrence of embedded atoms raises important questions, including the kinetic pathway leading to their formation. As mentioned in the introduction, there have been *theoretical* studies of Co atomic-wire formation at Cu(111) steps which can serve as useful points of comparison. One particularly detailed investigation, by Mo *et al.* [11], used DFT to study one- and two-atom wide wire phases of Fe, Co and W. In fact, the predictions of this study have been recently experimentally verified for the related case of Fe-wire growth, which was explicitly examined in [15]. For the case of Co wire formation, the theoretical study by Mo *et al.* determined a mechanism consisting of three steps: 1) formation of a single-atom-wide Co wire located one row behind the Cu step edge, 2) formation of a subsequent Co row behind the first, and 3) formation of a Co wire on top of these two rows. Despite being validated for the related case of Fe-wire growth, the predicted surface phase determined by this DFT-based calculation appears different from the phase present in our experimental results; specifically, non-encapsulated one and two-atom-wide wires are observed as opposed to the encapsulated wires predicted by calculation. Note however there are some differences between this theoretical study and the present study. For example, the calculations were performed for an A-type step, as opposed to a B-type step used experimentally. In addition, the calculations considered only a single Co atom at the step edge; higher coverages, i.e. a Co wire, might exhibit different energies. We note that the formation of the structures seen here at RT may be kinetically limited. For example, we observe a decrease in the Co Auger signal as the surface is raised above room temperature. We attribute this to either Co encapsulation by Cu or Co dissolution into the bulk. Thus higher temperatures may be required to reach the kinetic pathways leading to the final state of the DFT prediction, i.e. lateral step encapsulation. For the different scenario of deposition at a higher substrate temperature (instead of raising the substrate temperature post deposition), overall interdiffusion would probably dominate, as already evidenced by the presence of embedded terrace nucleation (Fig. 2(d)), rather than simply forming uniform laterally encapsulated wires. Finally, an additional comment on the lack of laterally encapsulated Co wires is discussed below.

An earlier theoretical study by Gomez *et al.* [10] using Monte Carlo and static relaxation provides a second interesting comparison of Co growth on Cu(111). In part, this study looked at preferential nucleation sites of Co at Cu steps and drew the following conclusions: (1) Co atoms will form "two nearly straight rows", one in front of the Cu step edge and the other behind the first row of Cu at the step edge (the latter was predicted later again by Mo *et al.*), (2) Co atoms were not likely to site exchange with Cu atoms on the terraces. While our work finds agreement with wire formation in front of the Cu step edge, there is no evidence of laterally encapsulated wire formation, as discussed above. Instead, the simultaneous presence of embedded-terrace nucleation suggests, paradoxically (see below), a lower energy of formation for this morphology than that of encapsulation. It should also be noted that the work of Gomez *et al.* did not stipulate the step type on which the simulation was performed.

In light of the two theoretical studies discussed above, our observation of terrace-site exchange without the presence of step-edge lateral encapsulation will be commented on briefly. One obvious consideration is that inclusion of additional physics appears to be important as is mentioned above. For example, the presence of significant "frizz" at the step edges indicates that Cu-atom diffusion and kink motion are present on theses edges. The presence of these surface defects and dynamics would seem to make calculation of the actual surface more complex than in the ideal theoretical models considered thus far.

In general, our observations indicate a degree of lack of uniformity in wire formation across the full surface at higher coverage. For example, as shown in Fig. 4(i), at 0.12ML, terrace nucleation, bi-atomic wires and bare steps can all be observed in adjacent regions of the STM image. In Fig. 4(ii) we show the change in the ratio of wire length to total step-edge length with coverage. These measurements show that while one-atom-wire growth dominates for 0-0.09ML and follows an ideal wire-growth pattern, the onset of two-atom-wire growth and terrace nucleation causes a reduction in the slope of one-atom wire concentration versus coverage. This change suggests that it is more energetically favorable for Co atoms to be attracted to other Co step-edge atoms, than to Cu step-edge atoms.

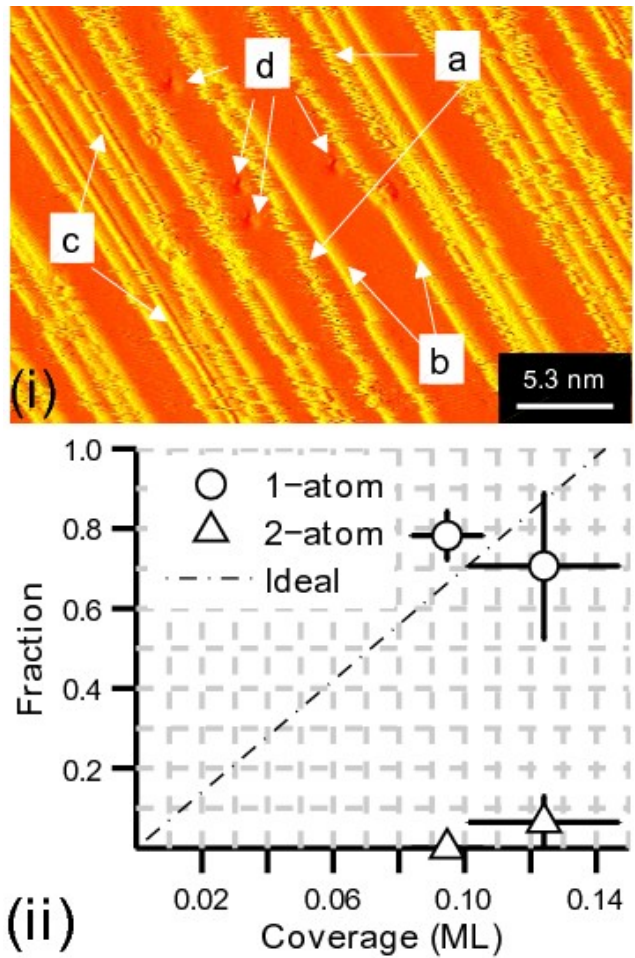

FIG. 4. (Color Online) Different Co growth phases develop with increasing coverage. Top panel (i): A derivative image shows that at 0.12ML, four different "phases" exist: (a) "frizz", (b) one atom-wide wire, (c) two atom-wide wire, (d) terrace embedded Co. See details in Fig. 2(b) for the single-atom-wide wires. Bottom panel (ii): Wire-length-to-step-length ratio versus coverage. "Ideal" refers to calculated ratio for growth of only 1-atom wide wires on a perfect seven-atom-row stepped surface. The falloff in the 1-atom wire ratio above 0.09 ML is due to 2-atom-wide wire growth, terrace nucleation, and terrace-width variations; without these components, the 1-atom wire step ratio would follow that of ideal wire growth.

The observed distribution of Co atoms along the Cu steps, i.e. extended Co wires coexisting with equally extended bare Cu steps, as shown in Fig. 5, allows us to speculate about the relative energetics of Co atoms on vicinal Cu surfaces. Judging from our measurement of the lengths of bare Cu steps at 0.09ML, Co atoms are sufficiently mobile on Cu (111) terraces even at 160 K, and the attractive potential of these atoms to the step edges is sufficiently shallow, that Co atoms may thermally migrate along the step edges at least 3.4 nm (our typical measured length of a clean Cu step edge at 0.09ML) before nucleating into an atomic wire. This is also evidenced by the random positions of the self-assembled wires along the step edge; in the case of Co on vicinal Au[25], on the other hand, Co atoms were found to nucleate at discommensuration lines that ran perpendicular to the steps, forming an ordered arrangement of quantum dots rather than extended atomic-wide wires. Once a wire nucleation on a Cu step edge has occurred, the ends of the wire serve as traps for otherwise mobile Co atoms. This attractiveness of the Co/Cu kinks at the ends of the wires is consistent with the growth of the extended uninterrupted Co wires.

The one-atom-wide wire-length distribution, as shown in Fig. 3, may be used to estimate the Co-Co interaction energy $\varepsilon$. We have used a one-dimensional lattice gas model [26] to fit the measured wire-length distribution, as shown by the gray curves in Fig. 3. The fit of the model to our wire-length distribution is particularly satisfactory, especially for the case of 0.09ML, at which coverage only 1-atom-wide wires are present. For the case of 0.12ML, 2-atom-wide wires also exist. This additional feature is not part of the model and thus the fit in that case is not as close.

In this lattice gas model, Yilmaz *et al.* derive an expression for the number of wires with length $l$, $q_l$, as

$$q_l = \frac{q^2}{n_1}\left(1 - \frac{q}{n_1}\right)^{l-1} \tag{1}$$

where $q$ is the number of wires and $n_1$ is the number of occupied sites along the step edge. By data fitting, we find $q$, and by the thermal equilibrium expression for $q$ (equation 5 in [26]), it is easily found that the interaction energy $\varepsilon$ can be written as

$$\varepsilon = k_B T \ln\left(\frac{(n-n_1)n_1 - nq + q^2}{q^2}\right) \tag{2}$$

where $n$ is the number of lattice sites along the step edge. Using the above formulation along with our measurements, we find an attractive Co-Co interaction energy of +5.1(±0.3) $k_BT$, or 0.13(±0.01) eV at RT. This energy is considerably lower than the bond energy expected for an isolated Co dimer[27]. This difference in bond energy between metal-substrate-supported wires and isolated wires is typical for wires on other surfaces. Thus, in the case of Ag on Pt(111), the Ag dimer bond energy has been experimentally found to be 0.150(±0.020) eV [28]. In addition, a comparable value of 0.166eV has been derived from STM measurements of 1-atom wide Ag wires on Pt(997)[29]. By contrast, the experimentally determined bond energy for isolated $Ag_2$ has been found to be 1.65(±0.03) eV [27]. In summary, the energy value extracted from the lattice gas fitting seems reasonable based on the observations of other atomic-wire systems.

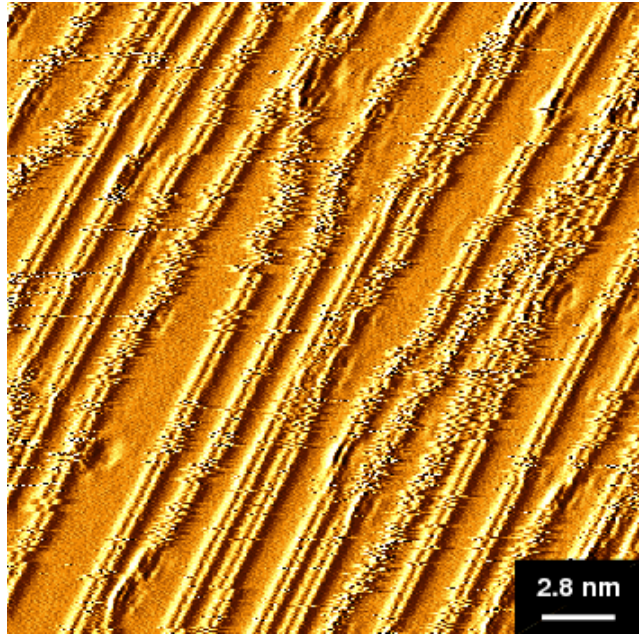

FIG. 5. (Color Online) Self-assembly at 0.09ML Co coverage. This STM image ($d^2z/dx^2$, i.e. second-derivative), acquired at RT, clearly shows the presence of self-assembled 1-atom-wide Co wires along with bare Cu(111) steps. The 1-atom-wide wires appear as straight sharp double lines at the step edge. Frizz, on the other hand, causes the bare Cu(111) step edge to appear irregular. The positions of the self-assembled wires are relatively random, indicating freedom from any substrate-mediated reconstruction or discommensuration. Note: the appearance of broken/shifted wires is merely due to tip-change.

Note that second-row Co nucleation is not observed before the Co coverage reaches ~0.12ML (see Fig. 4(ii)), i.e. before a significant fraction of the Cu steps are decorated by single-atom-wide Co wires. This observation suggests that Co atoms are sufficiently mobile also along the chains of Co atoms under experimental conditions. The latter property of the system is also important for the atomic wire growth as it prevents Co island formation at low Co coverages. Finally, it would be interesting to study the influence of the Co wires on the step fluctuations of bare Cu steps, i.e. "frizz"; such a study would provide additional insight into any change in step-edge transport, see for example Tao *et al.*[30].

More generally, the behavior of Co atoms on the vicinal Cu surface is strongly dependent on the temperature of the system. At temperatures higher than 160K the in-terrace diffusion of Co atoms becomes possible, as has been observed in our experiments at room temperature deposition (not shown in this work). At low temperatures, shorter wires and/or island growth would be expected, as in the case of Co on vicinal Pt[2], due to decreased mobility of Co atoms. On the other hand, the wire nucleation probability strongly depends on the concentration of mobile Co atoms on the surface. This concentration was controlled by the Co deposition rates in our experiments. The higher deposition rates would lead to higher nucleation probability and, therefore, shorter wire lengths. Thus, our work offers the empirically found set of conditions (160K, 0.01ML/min), suitable for growth of extended Co wires on a Cu(775) surface. We also speculate that the coverage-limit of 0.09ML for the self-assembly of uniform one-atom wide Co wires may be increased under more controlled growth conditions.

Straight sharp-edge wires are only seen at very low coverages, i.e. <0.2ML. If the coverage is increased above this value, a phase change is observed in the system, such that step bunching occurs and step edges appear irregular. This more complex growth morphology causes some terraces to become wider than others and as the coverage increases further, typically >0.3ML, island formation begins to occur. Similar higher-coverage effects have been reported by Gambardella *et al.* [2] for Co/Pt(997). A study of this higher coverage regime is currently being carried out.

The formation of one and two-atom wide Co wires, as well as terrace nucleation, is expected to modify the surface electronic structure of bare Cu(775). A recent photoemission investigation studied the modification of the Cu(775) surface state as a function of Co coverage [3]. The study clearly showed that the parallel momentum corresponding to the band minimum of the Cu surface state shifts in position at 0.03ML coverage indicating a change from a surface modulated state to a terrace modulated state at coverages corresponding to the onset of one-atom-wide wire growth. Such a shift can be expected as the randomly distributed Co wires disturb the coherence across the step superlattice and thereby preventing a coherent surface state formation extending over several terraces. The state can however still exist within the confines of an individual terrace.

## V. Conclusion

In conclusion, we have used STM to study the formation of a new phase of one and two-atom-wide Co wires on vicinal Cu(111). These two types of wires form sequentially with increasing coverage. For this new phase, Co wires assemble only at the bottom or lower terrace of a step edge; this result contrasts with top-only or top-bottom step edge Co island growth, as seen in past experimental observations of low-coverage Co deposition on Cu(111). Interestingly, we find no evidence for lateral Cu encapsulation of the Co wires in the terrace at RT; this is in comparison to the different atomic-wire phases reported in recent theoretical works. Single-atom-wide wires are seen for coverages less than 0.09ML, but at higher coverage two-atom wide wires form; in addition terrace substitution is found to co-exist. An examination of the length distribution for low and high coverage is consistent with growth kinetics which allow high mobility of Co atoms along the step edges and attachment at wire ends. These results, obtained on a vicinal surface with terraces of 14Å average width, add new insight to the already rich set of self-assembly physics of the Co/Cu(111) bimetallic system; further theoretical study of energetics and kinetics would provide additional insight.

**Acknowledgements**

This research was supported by the Department of Energy Contract No. DE-FG 02-04-ER-46157. Work at Brookhaven National Laboratory was supported by the Department of Energy under Contract No. DE-AC02-98CH10886. We thank Mehmet Yilmaz, Jerry Dadap, and Cyrus Hirjibehedin for several useful comments and suggestions. We also thank our unknown reviewers for bringing our attention to the work of Gomez *et al.* [10].

[1]     D. Mattis, *The Many-Body Problem* (World Scientific, Singapore, 1993).

[2]     P. Gambardella, M. Blanc, L. Bürgi, K. Kuhnke, and K. Kern, Surf. Sci. **449**, 93 (2000).

[3]     S. Wang, M. B. Yilmaz, K. R. Knox, N. Zaki, J. I. Dadap, T. Valla, P. D. Johnson, and R. M. Osgood, Phys. Rev. B. **77**, 115448 (2008).

[4]     K. Ogawa, K. Nakanishi, and H. Namba, Surf. Sci. **566-568**, 406 (2004).

[5]     J. de la Figuera, J. E. Prieto, C. Ocal, and R. Miranda, Surf. Sci. **307-309**, 538 (1994).

[6]     J. de la Figuera, M. A. Huertagarnica, J. E. Prieto, C. Ocal, and R. Miranda, Appl. Phys. Lett. **66**, 1006 (1995).

[7]     M. Ø. Pedersen, I. A. Bönicke, E. Lægsgaard, I. Stensgaard, A. Ruban, J. K. Norskøv, and F. Besenbacher, Surf. Sci. **387**, 86 (1997).

[8]     H. W. Chang, F. T. Yuan, Y. D. Yao, W. Y. Cheng, W. B. Su, C. S. Chang, C. W. Lee, and W. C. Cheng, J. Appl. Phys. **100**, 084304 (2006).

[9]     O. Pietzsch, S. Okatov, A. Kubetzka, M. Bode, S. Heinze, A. Lichtenstein, and R. Wiesendanger, Phys. Rev. Lett. **96**, 237203 (2006).

[10]    L. Gómez, C. Slutzky, J. Ferrón, J. de la Figuera, J. Camarero, A. L. Vazquez de Parga, J. J. de Miguel, and R. Miranda, Phys. Rev. Lett. **84**, 4397 (2000).

[11]    Y. Mo, K. Varga, E. Kaxiras, and Z. Y. Zhang, Phys. Rev. Lett. **94**, 155503 (2005).

[12]   J. de la Figuera, J. E. Prieto, C. Ocal, and R. Miranda, Phys. Rev. B. **47**, 13043 (1993).

[13]   J. E. Prieto, J. de la Figuera, and R. Miranda, Phys. Rev. B. **62**, 2126 (2000).

[14]   S. Speller, S. Degroote, J. Dekoster, G. Langouche, J. E. Ortega, and A. Närmann, Surf. Sci. **405**, L542 (1998).

[15]   J. D. Guo, Y. Mo, E. Kaxiras, Z. Y. Zhang, and H. H. Weitering, Phys. Rev. B. **73**, 193405 (2006).

[16]   J. Frohn, M. Giesen, M. Poensgen, J. F. Wolf, and H. Ibach, Phys. Rev. Lett. **67**, 3543 (1991).

[17]   M. Giesen, Prog. Surf. Sci. **68**, 1 (2001).

[18]   E. Lundgren, B. Stanka, G. Leonardelli, M. Schmid, and P. Varga, Phys. Rev. Lett. **82**, 5068 (1999).

[19]   E. Lundgren, B. Stanka, M. Schmid, and P. Varga, Phys. Rev. B. **62**, 2843 (2000).

[20]   S. Fölsch, P. Hyldgaard, R. Koch, and K. H. Ploog, Phys. Rev. Lett. **92**, 056803 (2004).

[21]   F. Nouvertné, U. May, M. Bamming, A. Rampe, U. Korte, G. Güntherodt, R. Pentcheva, and M. Scheffler, Phys. Rev. B. **60**, 14382 (1999).

[22]   N. Knorr, M. A. Schneider, L. Diekhöner, P. Wahl, and K. Kern, Phys. Rev. Lett. **88**, 096804 (2002).

[23]   J. A. Stroscio and R. J. Celotta, Science **306**, 242 (2004).

[24]   H. C. Manoharan, C. P. Lutz, and D. M. Eigler, Nature **403**, 512 (2000).

[25]   V. Repain, J. M. Berroir, S. Rousset, and J. Lecoeur, Surf. Sci. **447**, L152 (2000).

[26]   M. B. Yilmaz and F. M. Zimmermann, Phys. Rev. E. **71**, 026127 (2005).

[27]   M. D. Morse, Chem. Rev. **86**, 1049 (1986).

[28]   H. Brune, G. S. Bales, J. Jacobsen, C. Boragno, and K. Kern, Phys. Rev. B. **60**, 5991 (1999).

[29]   P. Gambardella, H. Brune, K. Kern, and V. I. Marchenko, Phys. Rev. B. **73**, 245425 (2006).

[30]   C. Tao, T. J. Stasevich, T. L. Einstein, and E. D. Williams, Phys. Rev. B. **73**, 125436 (2006).